\documentstyle[11pt] {article}

\title{Perturbative Analysis of Possible Failures in the Traditional Adiabatic Conditions}
\author{  T. V\'ertesi$^1$ and R. Englman$^{2}$ \\
$^1$  Section of Theoretical Physics, Institute of Nuclear Research\\
of the Hungarian Academy of Sciences, Debrecen, Hungary \\$^2$ Soreq NRC,~ Yavne ~ 81800,~Israel\\
 e-mails:tvertesi@dtp.atomki.hu;~englman@vms.huji.ac.il}

\begin{document}
\maketitle
\newcommand{\beq} {\begin{equation}}
\newcommand{\enq} {\end{equation}}
\newcommand{\ber} {\begin{eqnarray}}
\newcommand{\enr} {\end{eqnarray}}


\begin{abstract}

Recently, Marzlin and Sanders (2004) demonstrated an inconsistency when the
adiabatic approximation was applied to specific, "inverse" time-evolving
systems. Following that, Tong et al. (2005) showed that the widely used
traditional adiabatic conditions are insufficient to guarantee the validity
of the adiabatic approximation for this class of systems. In this article we
explore the origin of these observations by a perturbative approach and find
that in first order approximation certain nonzero terms appear in the
solution which gives rise to the breakdown of the adiabatic approximation
(despite the fact that the traditional adiabatic conditions are satisfied).
We argue that in this case the Hamiltonian of Marzlin and Sanders' inverse
time evolving system cannot be written in terms of $t/T$, where $T$ denotes
the total evolution time. It is further demonstrated that the new
qualitative adiabatic condition of Ye et al. (2005) performs well in some
cases when the traditional conditions fail to describe properly
non-adiabatic evolution.

\end{abstract}

\section{Introduction}

The quantum adiabatic theorem is one of the oldest fundamental results in
quantum physics \cite{Born}. It concerns development of systems where the
nondegenerate Hamiltonian evolves slowly in time. In the limit when the
change of Hamiltonian $H(t)$ is made infinitely slow the system, which
started from one of the eigenstates of $H(0)$, passes through the
corresponding instantaneous eigenstate of $H(t)$
\cite{Kato,Messiah,Avron,Ambainis}. The adiabatic theorem underlies the
adiabatic approximation scheme, which states that if the Hamiltonian $H(t)$
evolves slowly enough by satisfying the adiabatic condition in time interval
$t\in[0,T]$, then the evolving state of the system will remain close to its
instantaneous eigenstate up to a multiplicative phase factor in the interval
$[0,T]$. The adiabatic approximation has potential applications in several
areas of physics such as the Landau-Zener transition in molecular physics
\cite{LZ}, Gell-Mann-Low theorem in quantum field theory \cite{Gell-Mann} or
in the lore of Berry phase \cite{Berry}. Recently, with the emergence of the
new field of quantum information theory , new interest has arisen in the
application of the quantum adiabatic theorem \cite{Nielsen}. An alternative
scheme appeared beside the usual quantum algorithms, based on an adiabatic
time evolution of the state of the Hamiltonian where the final Hamiltonian
encodes the solution for a given problem \cite{Farhi}. In another
application, the Berry phase has been proposed to perform quantum
information processing tasks \cite{Jones}. Both of these concepts exploit
the circumstance, that the adiabatically evolving ground state is very
robust against decoherence and small perturbations \cite{Palma}. It is thus
important to explore the limits of the adiabatic conditions in order to
understand better when the evolution of a quantum system can be considered
adiabatic.

In a recent Letter Marzlin and Sanders \cite{Marzlin} demonstrated on a dual
pair of systems (called a-system and b-system), that if the evolution
operator of the b-system is the Hermitian conjugate of an adiabatically
evolving a-system, then the application of the adiabatic approximation in
the b-system can lead to contradiction. Tong et al.~\cite{Tong05} explained
this inconsistency by pointing out that the widely used traditional
adiabatic conditions are not sufficient to guarantee adiabaticity.

In the present paper our aim is to find the common root of the problems by
analyzing the time evolution of the dual systems in a perturbative manner.
In a first order approximation we obtain certain non-vanishing terms in the
regime where the traditional adiabatic conditions hold. We point out that
these nonzero terms are responsible for the violation of the adiabatic
approximation. We further argue that in this case the Hamiltonian of the
b-system cannot be written in terms of $t/T$, where $T$ denotes the total
evolution time. In addition, an explicit example is provided which on one
hand illustrates our arguments, and on the other hand allows a new adiabatic
condition of Ye et al.~\cite{Ye} to be tested on.

\section{Analysis of dual pair of systems}
\subsection{Expansion of state}

Consider a closed $N$-dimensional quantum system in a state
$|\psi(t)\rangle$, which evolves through the time-dependent Schr\"odinger
equation
\begin{equation}
i\frac{d}{dt}|\psi(t)\rangle=H(t)|\psi(t)\rangle\;, \label{sch}
\end{equation}
where $H(t)$ denotes the time-dependent non-degenerate Hamiltonian of the
system and we set $\hbar=1$. Let us introduce the normalized time $s$ by the
variable-transformation $t=sT$, $0\le s\le 1$, and rewrite equation
(\ref{sch}) as
\begin{equation}i\frac{\partial}{\partial s}|\psi(s,T)\rangle=T H(s,T)|\psi(s,T)\rangle\;,
\label{schs} \end{equation} where $T$ denotes the total evolution time. The
instantaneous eigenstates $|E_n(s,T)\rangle$ of the Hamiltonian $H(s,T)$
satisfy
\begin{equation}H(s,T)|E_n(s,T)\rangle=E_n(s,T)|E_n(s,T)\rangle\;,\quad n=1,\ldots,N\;,
\label{eigen}
\end{equation}
and the elements of the non-adiabatic coupling matrix are defined by
\begin{eqnarray}
\tau_{nk}(s,T)=\langle E_k(s,T)|\frac{\partial}{\partial s}
E_n(s,T)\rangle\;. \label{tau}
\end{eqnarray}
We now expand the state $|\psi(s,T)\rangle$ in the basis of the
instantaneous eigenstates of $H(s,T)$,
\begin{equation}
\psi(s,T)=\sum_{n=1}^N{\phi_n(s,T)e^{-iT\int_0^s{E_n(s',T)ds'}}|E_n(s,T)\rangle}\;.
\label{expand}
\end{equation}
(We shall frequently neglect in the text, but not in the formulae, the
variables ($s$,$T$) from the expressions following from equations
(\ref{schs},\ref{eigen},\ref{tau}), and the possible ($s$,$T$)-dependence is
understood without denoting it.)

\subsection{Parallel transport}

Let us now introduce the following local phase change for the $n$'th
instantaneous eigenstate
\begin{equation}
|\tilde{E}_n(s,T)\rangle=e^{i\Theta_n(s,T)}|E_n(s,T)\rangle\;,\quad
n=1,\ldots,N\;, \label{gaugetrafo}
\end{equation}
where $\Theta_n$ are real, ($s$,$T$)-dependent parameters. Plugging
(\ref{gaugetrafo}) into the definition (\ref{tau}) the transformation
formula for the non-adiabatic coupling terms \cite{Roi} reads
\begin{equation}
\tilde\tau_{nk}(s,T)=e^{i(\Theta_n(s,T)-\Theta_k(s,T))}\tau_{nk}(s,T)+i\frac{\partial\Theta_n(s,T)}{\partial
s}\delta_{nk}\;. \label{tautransform}
\end{equation}
This simple relation says that the diagonal element $\tilde{\tau}_{nn}$ is
boosted with respect to $\tau_{nn}$ by $i \partial\Theta_n/\partial s$,
while the non-diagonal elements of $\tilde{\tau}$ take up a phase with
respect to those of $\tau$. If we choose the phase
$\Theta_n(s,T)=i\int_0^s{\tau_{nn}(s',T)ds'}$, then under relation
(\ref{tautransform}) $\tilde{\tau}_{nn}$ becomes zero, and by definition
(\ref{tau}) the $n$th eigenstate satisfies the parallel transport law
$\langle E_n(s,T)|\partial/\partial s|E_n(s,T)\rangle=0$. Let us denote in
this gauge the matrix elements of $\tau$ by $\tau_{nk}^\|$, then we get
\begin{equation}
\tau_{nk}^\|(s,T)=e^{\int_0^s{\tau_{kk}(s',T)-\tau_{nn}(s',T)ds'}}\tau_{nk}(s,T)\;,
\label{tautoparallel}
\end{equation}
for $n\neq k$. Substituting $|\psi(s,T)\rangle$ expressed in the rotating
frame (\ref{expand}) into (\ref{schs}), and performing some algebra we
obtain for the complex amplitudes $\phi_n$ the following differential
equation:
\begin{equation}
\frac{\partial\phi_k(s)}{\partial s}=-\sum_{n\neq
k}\phi_n(s,T)\tau_{nk}^\|(s,T)e^{-iT \int_0^s{g_{nk}(s',T)ds'}}
\;, \label{diff}
\end{equation}
where $g_{nk}(s,T)\equiv E_n(s,T)-E_k(s,T)$. Now let us define the matrix
elements $A_{nk}(s,T)$ by
\begin{equation}
A_{nk}(s,T)=\frac{\tau_{nk}^\|(s,T)}{g_{nk}(s,T)}\;. \label{ank}
\end{equation}
The widely used, traditional condition for the adiabatic approximation, if
the system starts its evolution in the $n$'th instantaneous eigenstate of
the Hamiltonian $H(t)$, is then encoded in the statement
\cite{Messiah,Marzlin,Tong05,Aharonov,MacKenzie}
\begin{equation}
\left|\frac{\langle
E_k(t)|\dot{E}_n(t)\rangle}{E_k(t)-E_n(t)}\right| \ll 1 \;,\quad
k\neq n\;,\quad t\in[0,T]\;, \label{acontrad}
\end{equation}
which under the variable transformation $t=sT$ is equivalent to the
condition \cite{Sarandy}
\begin{equation}
\left|A_{nk}(s,T)\right| \ll T \;,\quad k\neq n\;,\quad
s\in[0,1]\;. \label{acon}
\end{equation}

\subsection {$a$ and $b$ systems}

Let us now turn our attention to a pair of $N$-dimensional quantum systems
\cite{Marzlin,Tong05,Duki}, where the a-system defined by Hamiltonian
$H^a(t)$ and its dual b-system with Hamiltonian $H^b(t)$ are related through
the following formula:
\begin{equation}
H^b(t)=-U^{a\dagger}(t)H^a(t)U^a(t)\;, \label{dual}
\end{equation}
where we assumed that the spectrum of $H^a(t)$ is entirely discrete and
non-degenerate. Formula (\ref{dual}) implies $U^b(t)=U^{a\dagger}(t)$
between the evolution operators of the a-system and the b-system, and links
the eigenvalues and the eigenvectors of the dual systems through the
equations $E_n^b(s,T)=-E_n^a(s,T)$ (and equivalently
$g_{nk}^b(s,T)=-g_{nk}^a(s,T)$) and
$|E_n^b(s,T)\rangle=U^{a\dagger}(s,T)|E_n^a(s,T)\rangle$. According to Duki
et al.~\cite{Duki} after some manipulation of these formulae one may arrive
at the following correspondence between the matrix elements of $\tau^\|$ for
the dual pair of systems:
\begin{equation}
\tau_{nk}^{\|b}(s,T)=\tau_{nk}^{\|a}(s,T)
e^{-iT\int_0^s{g_{nk}^a(s',T)ds'}}\;. \label{tauatob}
\end{equation}
It is noted that since according to the definition (\ref{dual}) the a-system
and the b-system are interchangeable, the equations numbered $\left(13 -
20\right)$
have dual pairs by exchanging the labels (a) and (b). Now let us write out
equation (\ref{diff}) specifically for the b-system,
\begin{equation}
\frac{\partial\phi_k^b(s)}{\partial s}=-\sum_{n\neq
k}\phi_n^b(s,T)\tau_{nk}^{\|b}(s,T)e^{-iT
\int_0^s{g_{nk}^b(s',T)ds'}} \;, \label{diffb}
\end{equation}
By use of (\ref{tauatob}) this results in a simpler expression for the
evolution equation of the b-system \cite{Duki},
\begin{equation}
\frac{\partial\phi_k^b(s,T)}{\partial s}=-\sum_{n\neq
k}\phi_n^b(s,T)\tau_{nk}^{\|a}(s,T) \;. \label{diffbsimple}
\end{equation}

\subsection{Perturbational solution}
In the following we consider the above pair of equations
(\ref{diffb},\ref{diffbsimple}) and solve each of them perturbatively. Let
us consider that the initial state of the b-system is $|E_n^b(0,T)\rangle$
with $\phi_k^b(0,T)=\delta_{nk}$. As zeroth order solution, the amplitudes
$\phi_k^b$ are assumed not to be evolving, i.e.
$\phi_k^b(s,T)\approx\phi_k^b(0,T)$, which constitutes the adiabatic
approximation, and in turn by insertion into the rhs of (\ref{diffb}) and
(\ref{diffbsimple}) yields for $\phi_k^b\;$ the parallel set of first order
solutions shown below. Performing an integration by parts as well, the first
order solution for (\ref{diffb}) is given by
\begin{equation}
\phi_k^b(s,T)=\phi_k^b(0,T)+\frac{i}{T}\sum_{n\neq
k}{\left(P_{nk}^b(s,T)+Q_{nk}^b(s,T)\right)} \;, \label{solb}
\end{equation}
where the terms $P_{nk}^b$ and $Q_{nk}^b$ are specified by
\begin{equation}
P_{nk}^b(s,T)=A_{nk}^b(0,T)-A_{nk}^b(s,T)e^{-iT\int_0^s{g_{nk}^b(s',T)ds'}}
\;, \label{pnk}
\end{equation}
and
\begin{equation}
Q_{nk}^b(s,T)=\int_0^s{e^{-iT\int_0^{s'}{g_{nk}^b(s'',T)ds''}}\frac{\partial
A_{nk}^b(s',T)}{\partial s'}ds'} \;, \label{qnk}
\end{equation}
for $n\neq k$. On the other hand, the first order solution of
(\ref{diffbsimple}) is
\begin{equation}
\phi_k^b(s,T)=\phi_k^b(0,T)+\sum_{n\neq
k}{\int_0^s{\tau_{nk}^{\|a}(s',T)ds'}} \;. \label{solbsimple}
\end{equation}
(The first term on the rhs of (\ref{solb}) and of (\ref{solbsimple})
corresponds to the zeroth order solution.) We note that equation
(\ref{solb}) would be formally identical to Eq.~(21) in ref.~\cite{Sarandy}
if Sarandy et al. fixed the gauge to obey the parallel-transport law
$\tau_{nn}=0$. However, the main difference lies in the fact that Sarandy et
al. do not allow $H^b(s,T)$ to depend on the total evolution time $T$. By
contrast,  in our treatment the Hamiltonian of the b-system and all the
other quantities derived from it accommodate a possible $T$-dependence. (It
is noted that from the notion of $T$-dependence we exclude the trivial
situation when $T$ appears as $T/T'$, where $T'$ is different from but the
same order as $T$, since here the presence of $T$ is immaterial.)

Now let us consider the case \cite{Sarandy,MacKenzie} when the Hamiltonian
of the a-system $H^a$ is a (continuous) function of $s$ alone. Then both
$\tau_{nk}^{\|a}$ and $E_n^a$ depend  only on $s$, which in turn entails
that $\partial A_{nk}^a(s)/\partial s$, appearing in the dual equation of
(\ref{qnk}), is a function of $s$ alone, as well. However, the phase of the
exponential in this equation does depend on $T$. Then it follows from the
Riemann-Lebesgue lemma \cite{R-L} that $Q_{nk}^a(s)$ goes to zero in the
limit $T\rightarrow\infty$. This is the argument of Sarandy et
al.~\cite{Sarandy}, which allows us to keep only the first $P_{nk}^a(s)$
terms under the summation in the dual equation of (\ref{solb}) and then uses
the adiabatic condition (\ref{acon}) for the a-system, which in turn
guarantees that the adiabatic approximation is accurate.

Next let us focus our attention on the Hamiltonian of the dual b-system
$H^b$, while  we still assume, as above, that $H^a$ is $T$-independent. One
can see through the relation (\ref{tauatob}) that $\tau_{nk}^{\|b}$ depends
on $T$ on a rapid oscillatory manner and through the relation (\ref{ank})
that this applies also to $\partial A_{nk}^b(s)/\partial s$. Namely, we
obtain after some algebra
\begin{equation}
\frac{\partial A_{nk}^b(s,T)}{\partial
s}=e^{-iT\int_0^s{g_{nk}^a(s')ds'}}\left(iT\tau_{nk}^{\|a}(s)-\frac{\left(\partial\tau_{nk}^{\|a}(s)/g_{nk}^a(s)\right)}{\partial
s}\right)\;, \label{dankds}
\end{equation}
where the notation $g_{nk}^a(s)$ and $\tau_{nk}^{\|a}(s)$ refer to their
$T$-independence. By substitution of this equation into definition
(\ref{qnk}), in the limit $T\rightarrow\infty$ we obtain
\begin{equation}
Q_{nk}^b(s,T)\simeq iT\int_0^s{\tau_{nk}^{\|a}(s')ds'}\;.
\label{qnkbapprox1}
\end{equation}
Thus, provided that $|\tau_{nk}^{\|a}|$ in the above equation is not the
constant zero function, $Q_{nk}^b(s,T)$ may go to infinity in the large $T$
limit. The reason that for $T\rightarrow\infty$, $Q_{nk}^b$ behaves
extremely differently than its counterpart $Q_{nk}^a$ (which rather tends to
zero), can be understood by the fact that (contrary to $\partial
A_{nk}^a/\partial s$)~ $\partial A_{nk}^b/\partial s$ is not a function of
$s$ alone, and moreover it depends on $T$ in a fast oscillatory manner (see
equation (\ref{dankds})). Hence, in the limit of $T\rightarrow\infty$,
$\partial A_{nk}^b/\partial s$ does not converge for any $s\in[0,1]$  and in
turn the Riemann-Lebesgue lemma is not applicable, unlike in the a-system.

\subsection{T-dependence of the b-Hamiltonian}

It is also possible to prove that when $H^a$ depends on $s$ only (as we
assumed above), then $H^b$ depends also on $T$, and consequently $H^b$ can
not be written in the only terms of $s=t/T$. The proof proceeds by reductio
ad absurdum. Suppose that $H^b$ is $T$-independent. Hence the eigenvalues
$E_n^b$ (or equivalently $g_{nk}^b$) and the eigenstates $|E_n^b\rangle$ of
$H^b$ are also independent on $T$. According to definition (\ref{tau}) the
$T$-independence also applies to $\tau_{nk}^b$, and by virtue of relation
(\ref{tautoparallel}) it applies to $\tau_{nk}^{\|b}$ as well. However, due
to definition (\ref{ank}) $A_{nk}^b$ and then $\partial A_{nk}^b/\partial s$
should be also $T$-independent, which is the contradiction since $\partial
A_{nk}^b/\partial s$ under (\ref{dankds}) depends on $T$.

\subsection{Extended case}

So far, we have examined the b-system in the large $T$ limit, with the
constraint that the Hamiltonian of its dual a-system is independent on $T$.
Now we turn to a more general situation, by imposing the following pair of
conditions on the a-system: 1. it satisfies the adiabatic condition
(\ref{acon}), 2. $|\tau_{nk}^{\|a}|$ is strictly greater than zero $\forall
s\in[0,1]$. As we can see, now the possible $T$-dependence of the a-system
is not ruled out. In the following we prove that if these conditions are
met, then $P_{nk}^b$ is much smaller than $Q_{nk}^b$ and then $P_{nk}^b$ can
be neglected in the first order approximate solution with respect to
$Q_{nk}^b$. Let us remark that this situation is opposite to the one,
discussed before in the $T$-independent a-system, when $Q_{nk}^a$ were
negligible in relation to $P_{nk}^a$. Consequently, let us analyze the
fulfillment of
\begin{equation}
\left|\frac{Q_{nk}^b(s,T)}{P_{nk}^b(s,T))}\right|\gg 1
\;.\label{qperpcond}
\end{equation}
Subtraction of (\ref{solb}) from (\ref{solbsimple}) and further algebra lead
just to the quantity
\begin{equation}
\left|\frac{Q_{nk}^b(s,T)}{P_{nk}^b(s,T)}\right|=\left|1+\frac{iT\int_0^s{\tau_{nk}^{\|a}(s',T)ds'}}{P_{nk}(s,T)}\right|\quad\;,
n\neq k \;. \label{qperp}
\end{equation}
The term $P_{nk}^b$ in (\ref{pnk}) can be further written with the aid of
(\ref{tauatob}) and with the use of $g_{nk}^b(s,T)=-g_{nk}^a(s,T)$, to
obtain
\begin{equation} P_{nk}^b(s,T)=-(A_{nk}^a(0,T)+A_{nk}^a(s,T)) \;.
\label{pnk2}
\end{equation}
Inserting this result back to equation (\ref{qperp}) entails the formula
\begin{equation}
\left|\frac{Q_{nk}^b(s,T)}{P_{nk}^b(s,T)}\right|=\left|1-\frac{T}{(A_{nk}^a(0,T)+A_{nk}^a(s,T))}i\int_0^s{\tau_{nk}^{\|a}(s',T)ds'}\right|
\;. \label{qperp2}
\end{equation}
Due to the pair of conditions prescribed for the a-system above, the usual
adiabatic conditions (\ref{acon}) are satisfied, implying
$\left|T/(A_{nk}^a(0)+A_{nk}(s))\right|\gg 1$ in formula (\ref{qperp2}) and
$\left|\tau_{nk}^{\|a}(s)\right|$ is strictly greater then zero $\forall
s\in[0,1]$, hence the rhs of (\ref{qperp2}) is large enough so that
inequality (\ref{qperpcond}) holds true and in turn $P_{nk}^b$ can be
neglected in (\ref{solb}). In this case subtracting again (\ref{solb}) from
(\ref{solbsimple}) yields
\begin{equation}
Q_{nk}^b(s,T)\simeq iT\int_0^s{\tau_{nk}^{\|a}(s',T)ds'}\;,
\label{qnkbapprox2}
\end{equation}
which formula can be considered as a $T$-dependent extension of
(\ref{qnkbapprox1}). Thus now we have two expressions for $Q_{nk}$: an exact
formula (\ref{qnk}) expressed by the quantities of the b-system, and an
approximate formula (\ref{qnkbapprox2}) defined in terms of the
non-adiabatic couplings of the a-system, where the approximation is valid
provided the a-system satisfies the adiabatic conditions (\ref{acon}) and
besides $\left|\tau_{nk}^{\|a}(s) \right|$ is not an extreme small quantity
$\forall s\in[0,1]$. It is easy to construct an a-system which meets the
above requirements. However, according to the approximate formula
(\ref{qnkbapprox2}), and the pair of constraints which guarantee the
accuracy of (\ref{qnkbapprox2}), $Q_{nk}/T$ is non-vanishing. This implies
that in the first order solution (\ref{solb}) the amplitudes $\phi_k^b$ are
not constant in time, and in turn the evolution of the dual b-system cannot
be considered adiabatic. Next let us examine, whether the traditional
adiabatic conditions, despite of the invalidity of the adiabatic
approximation, are fulfilled in this b-system. According to definition
(\ref{ank}), relation (\ref{tauatob}) and $g_{nk}^b=-g_{nk}^a$, we obtain
\begin{equation}
|A_{nk}^b(s,T)|=|A_{nk}^a(s,T)|\;. \label{ankatob}
\end{equation}
Thus the b-system satisfies the traditional adiabatic condition (\ref{acon})
only when the a-system satisfies it \cite{Tong05}.

We can summarize that although in the b-system (constructed by relation
(\ref{dual}) and by imposing the double conditions above) the standard
adiabatic conditions are satisfied, due to the nonzero values $Q_{nk}^b/T$
the state of the b-system is not evolving adiabatically. Therefore it
represents an example that the standard, traditional adiabatic conditions
may not guarantee adiabaticity (Tong et al. \cite{Tong05}). On the other
hand, we argued that the Hamiltonian of this b-system can not be written in
terms of $t/T$ quantities, unlike the case when the standard adiabatic
conditions are sufficient in guaranteeing adiabaticity
\cite{Sarandy,MacKenzie}.

\section{Illustration of the Theory}

In the following the general analysis of the previous section is supported
by a simple example to demonstrate that the non-vanishing $Q_{nk}^b/T$ terms
has a key role, in that the standard adiabatic conditions are not sufficient
to guarantee the validity of the adiabatic theorem. To this end imagine that
the a-system is a spin-half particle at rest at the origin in the presence
of a constant magnitude magnetic field rotating in a plane at constant
angular velocity $\omega$. This type of system was also considered by
refs.~\cite{Wu,Pati}. The corresponding effective Hamiltonian in the rest
frame is
\begin{eqnarray}
 H^a(t)=-\frac{\omega_0}{2} \left(
\begin{array}{cc}
   0 &  e^{-i\omega t} \\
   e^{i\omega t}  &  0
 \end{array} \right)\;,
\end{eqnarray}
where $\omega_0$ is defined by the magnetic moment of the spin and the
strength of the magnetic field. The corresponding eigenvalues are
$E_1^a(s,T)=-E_2^a(s,T)=\omega_0/2$. Identifying $T=2\pi/\omega$ and
switching  to normalized time $s$ by the transformation $t=sT$, we obtain
for the normalized eigenspinors
\begin{equation}
|E_1^a(s,T)=\frac{\sqrt{2}}{2} \left( \begin{array}{cc} e^{-i\pi s
}\\-e^{i\pi s }
\end{array} \right)\;,\quad
|E_2^a(s,T)=\frac{\sqrt{2}}{2} \left( \begin{array}{cc} e^{-i\pi s
}\\e^{i\pi s}
\end{array} \right)\;,
\label{eigenspec}
\end{equation}
so that neither the Hamiltonian $H^a$ nor the corresponding eigenvectors are
$T$-dependent. The local phases of the eigenstates has been chosen so that
they obey the parallel transport law ($\tau_{nn}^a(s,T)=0$ for $n=1,2$). In
this case by direct substitution of (\ref{eigenspec}) into definition
(\ref{tau}) $\tau_{21}^{\|a}=\tau_{21}^{a}=-i\pi$. Calculating the
Hamiltonian $H^b$ of the dual system from (\ref{dual}) to the first order in
the small quantity $\frac{\omega}{\omega_0}$ one finds
\begin{eqnarray}
 H^b(s,T)-\frac{\omega}{2}=\frac{\omega_0}{2} \left(
\begin{array}{cc}
   1 &  0 \\
   0  &  -1
 \end{array} \right)- \frac{\omega}{2}
  \left(
\begin{array}{cc}
   \cos\omega_0 Ts & i \sin\omega_0 Ts \\
   -i\sin\omega_0 Ts  &  -\cos\omega_0 Ts
 \end{array} \right)\;
\end{eqnarray}
Evidently $H^b$ has an irremovable $T$-dependence. Let us now treat the
$b$-system perturbatively, by the general method given above. We consider
that the dual b-system is initially in its instantaneous
$|E_1^b(0,T)\rangle$ eigenstate, and calculate the relevant $Q_{21}^b(s,T)$
term both exactly and approximately by the mean of equations (\ref{qnk}) and
(\ref{qnkbapprox2}), respectively. The approximate solution according to the
previous section is good if the adiabatic condition (\ref{acon}) holds for
the a-system and that $|\tau_{21}^{\|a}|$ is nonzero. In our two-level model
the absolute value of $\tau_{21}^{\|a}$ is $\pi$, whereas condition
(\ref{acon}) implies $\omega\ll \omega_0$. If $T$ is big enough so that
$\omega\ll \omega_0$ fulfills, then the exact solution is approximated
sufficiently well. However, in the concrete calculations both the exact and
approximate solutions yield $Q_{21}^b(s,T)/T=-\pi s$, which surprising
result is due to the fact that the above pair of conditions are in fact not
necessary for the approximation would be faithful (i.e. they are merely
sufficient conditions). Thus the formula $Q_{21}^b/T= -\pi s$ tells us that
regardless of the value of $T$, $Q_{21}^b/T$ may become relevant provided
$s$ is greater then zero.

Let us now calculate how faithful the adiabatic approximation is in this
case. The adiabatic approximation is acceptable in the interval $t\in[0,T]$
if the overlap
$|\langle\psi(t)|E_1(t)\rangle|^2=|\langle\psi(s,T)|E_1(s,T)\rangle|^2\simeq
1$ for $s\in[0,1]$ provided the state of the system is prepared initially in
the eigenstate $|E_1(0)\rangle$. However,
\begin{equation}|\langle\psi(s,T)|E_1(s,T)\rangle|^2=1-|\langle\psi(s,T)|E_2(s,T)\rangle|^2=1-|\phi_2(s,T)|^2\;,\end{equation}
where the first equality is valid for two-level systems, and the expansion
formula (\ref{expand}) was used to obtain the second equality. For our
concrete example, according to (\ref{solbsimple}) in first order
approximation the amplitude $\phi_2^b$ reads
\begin{equation}
\phi_2^b(s,T)=-\int_0^s{\tau_{21}^{\|a}(s',T)}ds'=i\pi s \;,
\label{chain}
\end{equation}
and by returning to the original time variable $t=\frac{2\pi s}{\omega}$ we
can write this further as
\begin{equation}|\langle\psi^b(t)|E_1^b(t)\rangle|^2= 1-\left(\frac{\omega
t}{2}\right)^2\;. \label{fidapprox}\end{equation} In our simple example
$\tau_{21}^{\|a}$ does not depend on $s$ and equation (\ref{diffbsimple})
can be easily solved (without applying perturbation theory) to obtain the
exact solution $\phi_2^b(s,T)=\sin (\pi s)=\sin(\frac{\omega t}{2})$, which
yields the exact fidelity
\begin{equation}|\psi^b(t)|E_1^b(t)|^2=1-\sin^2\frac{\omega t}{2}\label{fidexact}\end{equation} for our
particular example. If we compare (\ref{fidexact}) with (\ref{fidapprox}),
one can see that the approximate formula (\ref{fidapprox}) reproduces the
first term of the Taylor series expansion of the $sine$ function in the
exact formula (\ref{fidexact}). Performing higher order perturbation by
means of an iterative solution of equation (\ref{diffb}), the resulting
series will approach more and more accurately to the exact expression for
the overlap. This implies that in order to solve (\ref{diffb}) by
perturbation method one has to take into account at each iterative step
non-negligible off-diagonal terms (where in the first order the
corresponding term is $Q_{nk}^b$ for $n\neq k$). These terms when we go
beyond the first order approximation accumulate and according to the exact
formula (\ref{fidexact}) for the overlap (alternatively see Tong et
al.~\cite{Tong05}) for sufficiently large $t$ results in a great deviation
from the ideal overlap 1, pertaining to the pure adiabatic evolution.

In summary, this simple example shows that correct to first order it is
indeed  $Q_{nk}\;, (n=1,k=2)$ term in (\ref{solb}) that is responsible for
the breakdown of the adiabatic approximation, despite  the fulfillment of
the traditional adiabatic conditions.

\section{Test of the novel adiabatic condition}

Finally, we test a new quantitative adiabatic condition, proposed by Ye et
al.~\cite{Ye}. The new quantity, introduced in ref. \cite{Ye}, looks as
\begin{equation}
\tilde{A}_{nk}(s,T)=\frac{\tau^{\|}_{nk}(s,T)}{g_{nk}(s,T)-\frac{1}{T}\frac{d\arg
\tau_{nk}^{\|}(s,T)}{ds}}\;,
\end{equation}
and the new quantitative adiabatic condition is encoded in the requirement
\cite{Ye}
\begin{equation}
|\tilde{A}_{nk}(s,T)| \ll T\;,\quad k\neq n\;,\quad s\in[0,1]\;,
\label{aconnew}
\end{equation}
in order the adiabatic approximation  be faithful in the time range
$t\in[0,T]$. By substitution of the relations (\ref{tauatob}) and
$g_{nk}^b=-g_{nk}^a$ between the dual systems, the following simpler
condition is given for the b-system,
\begin{equation}
\left|\frac{\tilde{A}_{nk}^b(s,T)}{T}\right|=\left|\frac{\tau_{nk}^{\|b}(s,T)}{Tg_{nk}^b(s,T)-\frac{\partial
arg\tau_{nk}^{\|b}(s,T}{\partial s}}\right| =\left|
\frac{\tau^{\|a}_{nk}(s,T)}{\frac{\partial\arg
\tau_{nk}^{\|a}(s,T)}{\partial s}}\right| \ll 1 \;.
\label{aconnewsimple}
\end{equation}
Next the above condition is applied for our special two-level b-system,
discussed in the previous section. Since $\tau_{21}^{\|a}(s,T)=-i\pi$, its
argument remains constant in time, and therefore quantity
$|\tilde{A}_{nk}^b(s,T)/T|$ is infinite for any $s\in[0,1]$ and in turn
(\ref{aconnewsimple}) is not fulfilled. This implies that the novel
adiabatic condition of Ye et al.~\cite{Ye} indicates correctly that the
adiabatic approximation is violated in the b-system, whereas the traditional
adiabatic condition fails to detect it.

\section{Conclusion}

In conclusion we investigated through a first order perturbative solution
the state evolution of a dual pair of systems, which construction was
considered first by Marzlin and Sanders \cite{Marzlin}. The a-system is
assumed to satisfy the usual adiabatic criteria, and the b-system performs a
reversed time evolution with respect to its dual pair. In this case we argue
that in the b-system due to the fast oscillations of the off-diagonal
non-adiabatic coupling terms ($\tau_{nk}^{\|b}$), the terms $Q_{nk}^b$
defined by (\ref{qnk}) may become large and dominate in the first order
solution of (\ref{diffb}). This results in the breakdown of the adiabatic
approximation in the b-system, although the standard adiabatic conditions
are satisfied in it. Thus, we could find via perturbative treatment the
underlying reason why the traditional adiabatic conditions are insufficient
in the b-system. It is further argued that in this case the Hamiltonian of
the b-system cannot be written in the form of $H(t/T)$. The above findings
are illustrated on a simple two-level model as well, whereon a new adiabatic
condition was also tested and showed that in this particular case it
performs well.

\end{document}